\documentclass[iicol,sn-vancouver,Numbered]{sn-jnl}% Default with double column layout
\numbered
\usepackage{natbib}
\usepackage{graphicx}%
\usepackage{multirow}%
\usepackage{amsmath,amssymb,amsfonts}%
\usepackage{amsthm}%
\usepackage{mathrsfs}%
\usepackage[title]{appendix}%
\usepackage{xcolor}%
\usepackage{textcomp}%
\usepackage{manyfoot}%
\usepackage{booktabs}%
\usepackage{algorithm}%
\usepackage{algorithmicx}%
\usepackage{algpseudocode}%
\usepackage{listings}%
\usepackage[english]{babel}
\addto\extrasenglish{
  
}
\addto\extrasenglish{
  
}
\addto\extrasenglish{
  
}

\usepackage{atbegshi,picture}

\AtBeginShipoutNext{\AtBeginShipoutUpperLeft{%
  \put(\dimexpr\paperwidth-1cm\relax,-1.5cm){\makebox[0pt][r]{\framebox{final version published in \emph{Künstliche Intelligenz}, 2024: \url{https://doi.org/10.1007/s13218-023-00820-x}
}}}%
}}

\begin{document}

\title[Article Title]{Building an AI Support Tool for Real-time Ulcerative Colitis Diagnosis}

\author*[1]{\fnm{Bjørn Leth} \sur{Møller}}\email{bjm@di.ku.dk}
\equalcont{These authors contributed equally to this work.}

\author[2,3]{\fnm{Bobby Zhao Sheng} \sur{Lo}}\email{bobby.zhao.sheng.lo@regionh.dk}
\equalcont{These authors contributed equally to this work.}
\author[2,3]{\fnm{Johan} \sur{Burisch}}\email{johan.burisch@regionh.dk}
\author[2,3]{\fnm{Flemming} \sur{Bendtsen}}\email{flemming.bendtsen@regionh.dk}
\author[2,3]{\fnm{Ida} \sur{Vind}}\email{ida.vind@regionh.dk}
\author[1]{\fnm{Bulat} \sur{Ibragimov}}\email{bulat@di.ku.dk}
\author[1]{\fnm{Christian} \sur{Igel}}\email{igel@di.ku.dk}

\affil[1]{\orgdiv{Department of Computer Science}, \orgname{University of Copenhagen}, \country{Denmark}}

\affil[2]{\orgdiv{Gastrounit, Medical Section}, \orgname{Copenhagen University Hospital Hvidovre}, \state{Hvidovre}, \country{Denmark}}

\affil[3]{\orgdiv{Copenhagen IBD Center}, \orgname{Copenhagen University Hospital Hvidovre}, \state{Hvidovre}, \country{Denmark}}

\abstract{Ulcerative Colitis (UC) is a chronic inflammatory bowel disease decreasing life quality through symptoms such as bloody diarrhoea and abdominal pain.
Endoscopy is a cornerstone of diagnosis and monitoring of UC. The Mayo endoscopic subscore (MES) index is the standard for measuring UC severity during endoscopic evaluation. However, the MES is subject to high inter-observer variability leading to misdiagnosis and suboptimal treatment.
We propose using a machine-learning based MES classification system to support the endoscopic process and to mitigate the observer-variability. 
The system runs real-time in the clinic and  augments doctors' decision-making during the endoscopy.
This project report outlines the process of designing, creating and evaluating our system. We describe our initial evaluation, which is a combination of a standard non-clinical model test and a first clinical test of the system on a real patient.  
}

\maketitle

\section{Introduction}\label{sec:intro}
Ulcerative Colitis (UC) is a chronic, relapsing-remitting disease causing considerable reduction in quality of life \cite{taku2020}. UC is characterised by inflammation and ulcers in the colon and rectum. Symptoms include bloody diarrhoea, rectal bleeding and abdominal pain. Surgical removal of  the whole colon (colectomy) might be necessary in severe cases. Notably, the prevalence and incidence of UC are increasing worldwide~\cite{increasing_prevalence}.

Endoscopic evaluation is a cornerstone in disease monitoring and treatment guidance. 
However, the quality of the imagery can suffer from issues such as 
%It can suffer from various issues that affect the quality of the footage \cite{Good_endoscopy}. These complications typically result in noisy visuals, which can be attributed to factors such as 
motion blur, water interference, and the camera's contact with the colon wall \cite{Good_endoscopy}.
It is customary for clinicians to save specific video images captured during endoscopic evaluations. These images serve as valuable additions to a patient's medical record and are used as supportive evidence for the doctor's final disease severity evaluation.

\begin{figure*}
    \centering
    \includegraphics[width=\textwidth]{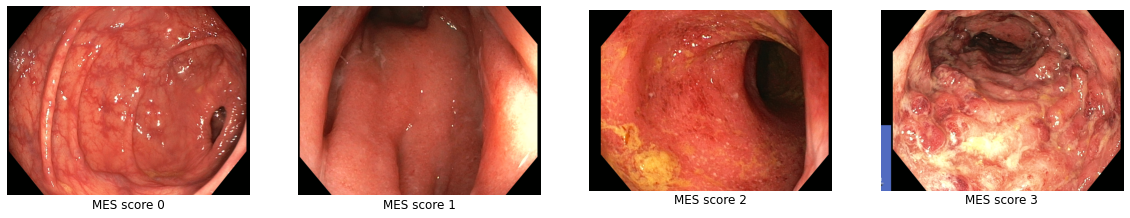}
    \caption{Examples of different endoscopic colon images and the corresponding MES scores }
    \label{fig:mes_score_examples}
\end{figure*}

The most common disease severity scoring index used during endoscopic evaluation is the Mayo endoscopic subscore (MES) which rates inflammation from 0 (remission) to 3 (severe inflammation) as seen in \autoref{fig:mes_score_examples}~\cite{schroeder1987}. 

The MES, however, is subject to both high inter-observer variability,  meaning that two different clinicians could disagree on the MES score of a frame, and intra-observer variability, meaning one doctor may assign the same frame different scores at different points in time. This bears the risk of impeding the efficacy of treatment plans for individual patients. Furthermore, misclassification can potentially result in the erroneous inclusion of patients in clinical trials, leading to wasted resources and weaker experimental results~\cite{wrong_treatment}. The inter-observer Cohen's kappa score, a common metric for measuring inter-observer variability, has been quantified to be as low as 0.6 among junior physicians, 0.7 among senior practitioners, and 0.8 when employing a centralized reading approach~\cite{bad_MES_stats}.

Using machine learning to assist with the analysis of endoscopic images and videos
%MES scoring has garnered significant 
is gaining attention in recent years \cite{ali2022we}. 
Recent work on machine learning for UC MES scoring is summarized in \cite{lo2021artificial}. 
In \cite{lo:22}, we trained a convolutional neural network (CNN) which achieved a test set accuracy of 0.84  in accurately distinguishing between all four MES scores. Notably, this performance level is comparable to that of senior gastroenterologists~\cite{bad_MES_stats}.
The majority of previous research, including our own, was primarily focused on classifying endoscopic images, manually taken and sorted by physicians, excluding noisy images.
In general, previous approaches would not transfer to the setting of raw (unfiltered) endoscopic footage due to the prevalence of noisy frames in these data.
Recently, some studies have addressed the challenge of analyzing endoscopic videos. Becker et al.~\cite{Benjamin_Gutierrez_Becker} leverage a two-stage modelling framework. They use one model for discarding poor-quality frames and another for MES scoring of prefiltered frames. 

Our goal is to create an automatic system that provides a second opinion during endoscopic procedures to increase diagnostic accuracy while ensuring that the clinician retains the ultimate responsibility for MES classification.
This requires us to parse raw endoscopic video footage in real-time while distinguishing between all MES classes. 
In an initially informal collaboration between the Department of Computer Science at  University of Copenhagen and the Gastro Unit, medical section, Copenhagen University Hospital Hvidovre, Denmark, we  explored machine learning-based automatic MES scoring to address the issue of inter-observer variability in the MES index \cite{Previous_work_on_dataset,lo:22}. The results  and the feedback from the medical community were encouraging, which motivated us to further develop the system to be used in  clinical practice. We received financial and organisational support from Hvidovre Hospital and a grant from BETA.HEALTH, which is an innovation platform supporting innovation projects at Danish hospitals with the primary objective to aid the development of new or improved treatments and clinical solutions 
This allowed us to  initiate the ENACT project (ENdoscopic Addon system for ulcerative Colitis patienTs). The aim of ENACT is to further develop our AI-based model into a solution designed to seamlessly integrate with existing endoscopic systems. This lowers the cost of adoption, because no changes to the core endoscopic system are needed. By eliciting requirements from experienced gastroenterologists, we formulated a system workflow that enhances the clinical decision-making process while ensuring minimal disruption.
The result of our efforts is HECTOR, a real-time system developed to aid clinicians during endoscopic evaluations, which we will present in the following.
At the heart of our system lies a machine learning-based model that has been trained to annotate images with MES classifications, emulating the collective judgment of a committee of clinical experts. Throughout the development of our system, we focus on providing value for clinicians, for example offering guidance to junior clinicians during endoscopic procedures.

In the following, we describe the system requirements of HECTOR, its current prototype implementation, and present the first clinical evaluation.

\section{System Requirements}\label{sec:req}

In close collaboration with clinical experts from Hvidovre Hospital, we identified the following requirements for the HECTOR system based on individual interviews and a 
user experience (UX) workshop involving ten highly experienced endoscopists:
\begin{enumerate}

    \item The system should distinguish between all four MES classes, not just between healthy (MES 0--1) and ill (MES 2--3) tissue.
    \item The system should provide an overall score for the endoscopic video and provide an explanation for the score.
    \item The system should parse raw endoscopic video footage. Thus, it needs to handle noisy data. 
    \item The system should run in real-time, so it can assist during the endoscopic evaluation and the second opinion of the model should be presented to the clinician in a non-distracting manner.
    \item The system should be simple to set up and use as it would  be employed in the time-constrained setting of a clinical hospital. Furthermore, the system should be hardware agnostic, allowing it to be connected to a variety of different endoscopic devices to allow for adaptation across the field.
    \item The system should serve as a supportive tool for clinicians, providing feedback on the need for a better viewpoint before capturing photos for medical journals.
    \item The system should ideally be able to improve over time as it is used. Errors in model predictions are unavoidable, and thus, a simple way of correcting model errors should be available to clinicians.  
\end{enumerate}
An important observation from our UX workshop has been that doctors expressed a preference against receiving in-depth explanations for model predictions during endoscopic evaluations. The doctors noted that additional information such as model prediction certainty, as expressed by the model output logits, would only serve as a distraction.  

\section{Implementation}\label{sec:impl}
This section describes the implementation of HECTOR. First an overview of the system and the interaction flow is given. After that, we elaborate on the different sub components.  
\subsection{Overview}
Based on the specified requirements, we have developed the overall design specifications for the system.
A core model parses raw  footage from the endoscopic camera and returns an MES prediction for the current camera footage. To account for potential noise in the footage, the model  distinguishes between views suitable for MES annotation and those unsuitable for annotation, as outlined in \autoref{Real-time Inference}. Consequently, the system only assigns scores to frames deemed suitable for annotation. Additionally, the model provides feedback to the end user by indicating whether the current view is suitable or not. This feedback serves as a reference for clinicians, enabling them to reposition the camera if necessary before making any judgments.

To determine an overall score for the entire video, individual frame scores are combined as described in \autoref{Generating Overall Video Score}. Frames that significantly contribute to the overall score are saved as they serve as explanatory evidence for the overall assessment.

The stored frames are presented to the clinician. If the clinician disagrees with a particular frame's score, they have the ability to update it and store it along with the frame. This updated information can then be used to retrain the model, leveraging the clinician's input. We are currently exploring different retraining procedures (see also \autoref{sec:discussion}).

The overall user interaction flow  is as follows: a button press initiates and stops the model. Once the model has been stopped, all relevant information pertaining to the parsed video stream is displayed on an external touch screen. This interface allows the clinician to make modifications to the model scores and save frames that they consider to be significant.

\subsection{Core Model}
The core model scores high-quality footage and discards low-quality frames. Furthermore, it generates an overall score for the video.  The real-time results of the model are presented in a non-intrusive manner, ensuring minimal disruption to the clinician's workflow or visual experience.

\subsubsection{Real-time Inference} \label{Real-time Inference}
Our approach to simultaneously discarding noisy frames and scoring good frames is based on casting the task as an open-set recognition (OSR) problem~\cite{vaze2022openset}.
Within the OSR framework, the objective is to classify whether a given sample belongs to the in-distribution, which is represented by the distribution of the training data. Conversely, samples that do not belong to the in-distribution are considered part of the out-distribution or open-set. For samples belonging to the in-distribution, the model assigns a score, whereas samples belonging to the out-distribution are discarded. One of the advantages of the OSR framework is that it eliminates the need for training samples representing the out-distribution class.

In our specific context, the in-distribution data are high-quality in-body endoscopic images, while any other footage was considered out of distribution. To establish the in-distribution, we  used a corpus of annotated high-quality endoscopic still images from \cite{Previous_work_on_dataset}. This corpus addresses the issue of inter-viewer-observability by annotations  determined through majority voting of three independent UC experts. Majority voting has been shown to reduce the inter-viewer observability \cite{GOTTLIEB2021710}. 

We have solved the OSR task following the approach by Vaze et al.~\cite{vaze2022openset}. A closed-set classifier providing probabilistic output in the form of unnormalized scores (logits) is applied to each frame, and the  output logits are stored. The frame is classified to belong to the in-distribution by thresholding the maximum logit. If the frame is deemed above the threshold, the model logits are used to determine the MES score. The detection of out-distribution samples has been shown to be correlated with the closed-set classifier ability to correctly classify in-distribution samples \cite{vaze2022openset}. Thus, we should be able to improve the removal of bad frames indirectly by improving the scoring of MES frames.  

To enhance overall performance, we incorporate pre- and post-processing steps. To improve frame discarding and increase frame throughput, we  implemented a Laplacian-based blur filter and a colour ratio filter to discard obvious low-quality frames. Furthermore, to stabilize results across the temporal dimension, we applied a rolling mean filter to the model output.

The closed-set model was based on the 3.7m parameter ConVNeXtv2 atto architecture \cite{Woo2023ConvNeXtV2}. The model was pretrained on Imagenet \cite{deng2009imagenet} and fine-tuned on the data corpus from \cite{Previous_work_on_dataset}. We selected  a rather small model to increase throughput to meet real-time constraints. The model was calibrated using temperature scaling on a held-out validation dataset \cite{temperature}. 

\subsubsection{Generating Overall Video Score} \label{Generating Overall Video Score}
After evaluating the real-time model, the video score is determined by selecting the maximum score generated. Due to the temporal filter employed by the real-time model, the score is determined by the highest scored section of frames rather than by a single frame. This approach ensures a more accurate representation of the overall video score, taking into account the collective assessment of the model across consecutive frames or sections. This approach to video scoring was implemented after a discussion with our collaborating medical experts.

\subsubsection{Displaying Real-time Results}
Following an end-user test, we decided to present the real-time model results on a picture-in-picture screen, ensuring no direct interference with the endoscopic footage displayed to the doctor. Additionally, we opted not to include any additional information, such as model confidence. These choices were based on feedback from the doctors, who specifically requested a simple and non-distracting interface with only the most important information (i.e., the MES classification).

\subsection{Human System  Interaction}
Once an endoscopy procedure has been completed, the clinician can interact with the system to obtain the score of the overall video. The system automatically selects frames that it deems to be interesting and of high quality, allowing the doctor to potentially include them in the medical journal of the patient.  Additionally, the doctor can correct  mistakes in the model predictions, thereby enabling the system to improve and refine its performance over time.

To provide this functionality, the model stores the $k$ most relevant frames during runtime. After the endoscopy procedure, these frames are presented to the clinician as an explanation for the model's overall video score. The clinician has the flexibility to select any number of the $k$ frames deemed important and add them to their medical journal. Moreover, the clinician has the ability to correct the scores associated with each frame in case of model errors. These corrected annotations are stored and can be used as additional data for the purpose of retraining the model, thereby allowing the system to adapt and improve over time.

The model's selection of the most important frames is determined by several factors, including the model's MES score, model certainty, and the distance between selected frames. Frames with higher MES scores are given priority over those with lower scores when comparing their relevance. Additionally, frames with higher model certainty, as indicated by probabilistic interpretation of the model output, are favoured. Lastly, a minimal distance between selected frames is enforced to ensure a diverse and representative frame selection.

This selection policy is motivated by the observation that our model exhibits higher confidence when scoring non-noisy observations. Therefore, the frames selected should predominantly consist of high-quality images suitable for inclusion in medical journals. Given that the overall video score is based on the highest scored MES sequence, the selection process is designed to prioritize high scoring frames, thereby allowing the clinician to correct any potential model errors effectively.

\subsection{Hardware}

\begin{figure}
    \centering
    \includegraphics[width=\columnwidth]{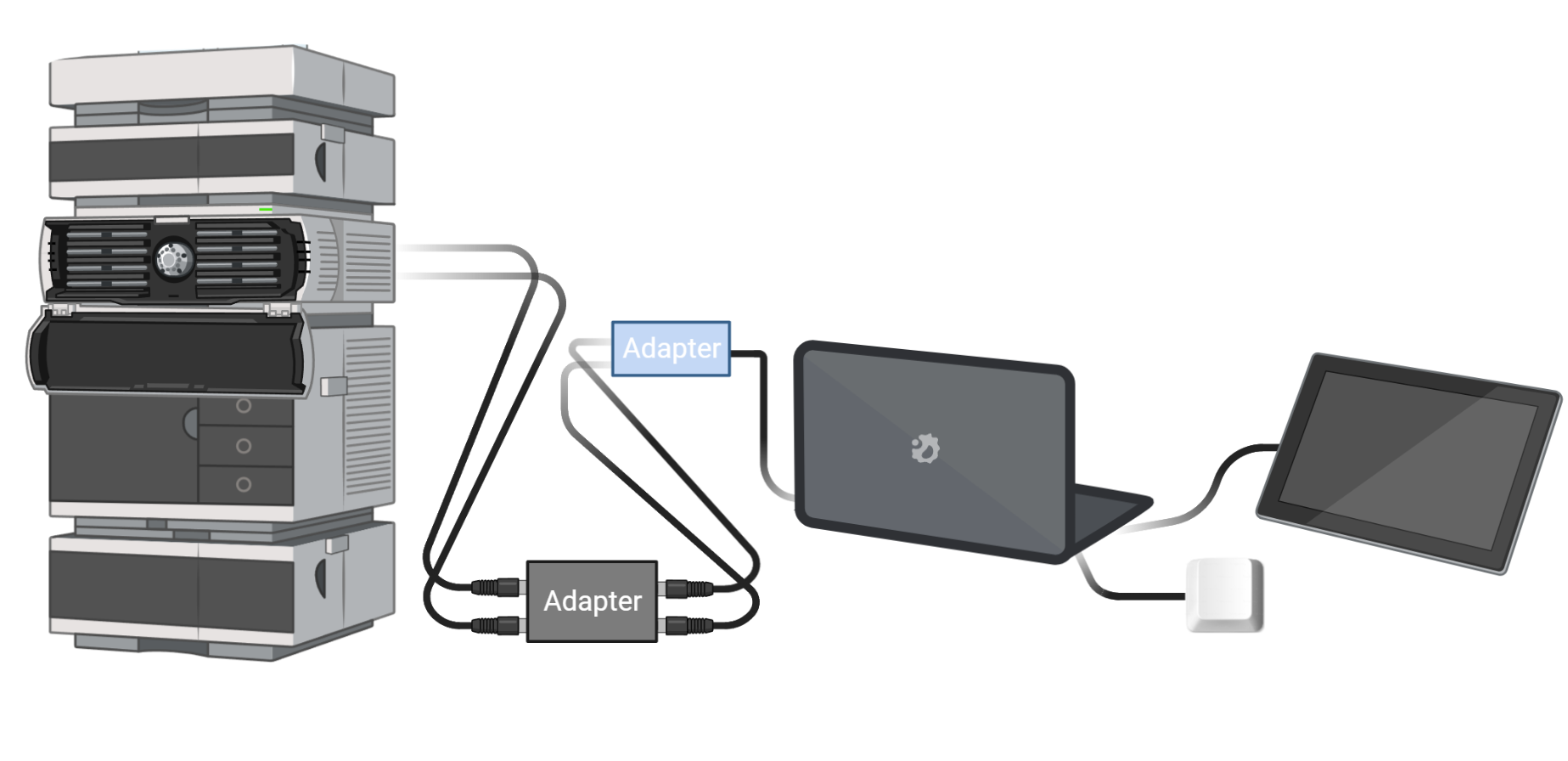}
    \caption{Graphical overview of the hardware setup}
    \label{fig:hardware}
\end{figure}

The hardware of the system is designed with ease of use for clinicians and seamless integration with various endoscopic devices in mind.
Our solution consists of two main components: a dedicated computer responsible for running the core system and an external touchscreen interface that enables clinicians to interact with the system following the endoscopy procedure. Additionally, a separate standalone button is used to start and stop the real-time system. A graphical overview can be seen in \autoref{fig:hardware}.
By employing an external computer instead of one integrated directly into the camera, our system can be  seamlessly integrated with a wide range of endoscopic systems that support multiple SDI inputs and outputs. 
The decision to incorporate a button as the means of interaction  with the system during the procedure is motivated by the need to provide doctors with a convenient way to interact with the system while simultaneously handling the endoscopic camera during procedures. 
% General Hardware setup.

\section{Evaluation}\label{sec:eval}
A comprehensive clinical evaluation of the system will require a sufficient amount of clinical data to be generated. While the collection of substantial clinical data is ongoing, we have obtained preliminary results that provide initial insights into the performance of the core model. To achieve an initial validation of our overall system, we conducted our first trial in a clinical setting involving a real patient. 
\subsection{Core Model Evaluation}
To approximate the performance of the core real-time model in a clinical scenario, we conducted two different non-clinical evaluations.

Firstly, to assess the model's ability to classify in-distribution into correct MES classes, we trained and evaluated the model using the corpus and training-validation split from \cite{Previous_work_on_dataset}. The model achieved a test accuracy of 0.84, corresponding to a Cohen's kappa of 0.79, indicating its capability to accurately classify images into the appropriate MES categories.

\begin{figure}
    \centering
    \includegraphics[width=\columnwidth]{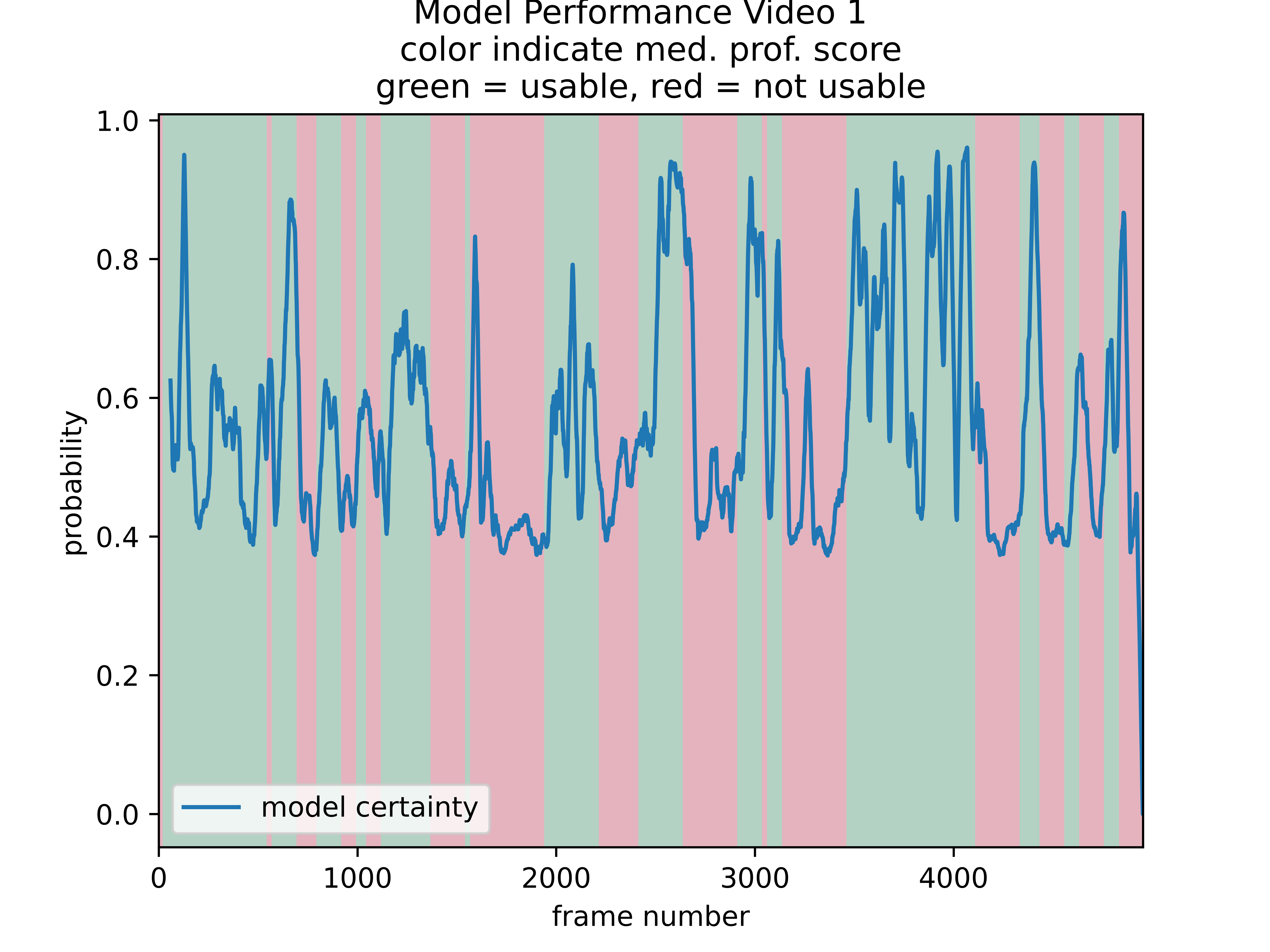}
    \caption{An example of using model logits to determine whether a frame is usable. Background colour displays whether the doctor deemed the frame usable or not. In general, there is a high agreement between the model certainty and the usability of the frame. }
    \label{fig:model_discard}
\end{figure}

Secondly, we evaluated the model's ability to filter raw video data into useful and non-useful frames. In collaboration with a medical doctor, we segmented nine videos into two categories: useful and non-useful frames. Subsequently, our system's video filtering capability was evaluated on this dataset, achieving a macro average AUROC of 0.904. The individual video AUROCs ranged between 0.975 and 0.77, demonstrating the effectiveness of our system in distinguishing between useful and non-useful frames. An example of a  segmented video  can be seen in \autoref{fig:model_discard}.
% Model results
\subsection{First Clinical Trial}
Our aim for the near future is to deploy the system in a range of clinics to generate diverse experimental data to explore if and how our system generates value to the clinicians. Before that, we need to ensure that the overall workflow can be integrated into  a clinical environment and that the  system is easy to operate by end-users. Thus, we  performed a first clinical trial using the system. Note that the main purpose of this test was not to quantify the model's accuracy but to test whether a clinician could use the system without assistance. 

We performed our first clinical trial of the system with a real patient and with the help of an experienced clinician. The goal was to validate the control flow and to gauge the initial value to a working clinician. The clinical trial was approved by the hospitals and the capitals regions' jurisdictional board. The patient was well informed of the test and procedure and consented to the test. The endoscopic hardware used during the procedure were Olympus EVIS EXERA III CV-190 PLUS and Olympus EVIS EXERA III GIF-1TH190. 

\begin{figure}
    \centering
    \includegraphics[width=\columnwidth]{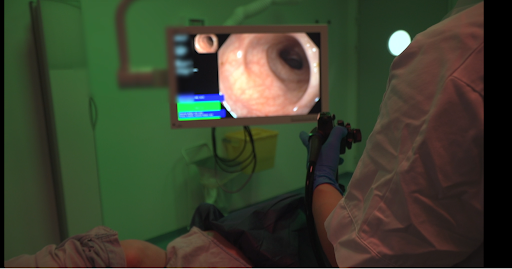}
    \caption{An example of our system scoring a view as MES 0 ( green colour in the button left)}
    \label{fig:model_0}
\end{figure}

\begin{figure}
    \centering
    \includegraphics[width=\columnwidth]{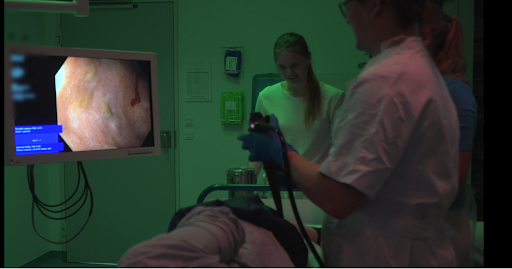}
    \caption{An example of our system scoring a  view as MES 1 ( blue colour in bottom left)}
    \label{fig:model_1}
\end{figure}

\begin{figure}
    \centering
    \includegraphics[width=\columnwidth]{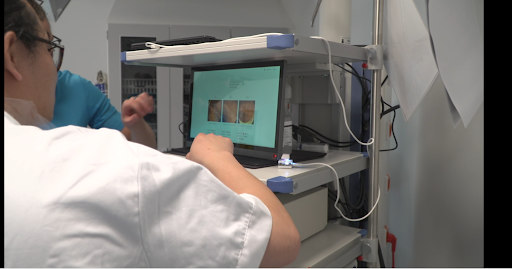}
    \caption{The screen where the doctor interacts with the system after the endoscopy }
    \label{fig:model_post}
\end{figure}

The results were positive. The interaction between the clinician and the system was without complications. No interventions from technicians were required during the operation and the clinician reported that the system was easy to use.
Images from the clinical trial can be seen in \autoref{fig:model_0}, \autoref{fig:model_1}, and \autoref{fig:model_post}. 
The MES scores of the system, on both frame and video level, were aligned with the scores of the clinician, giving an initial performance measure of the model. 
Of course, more trials need to be conducted to properly validate the control flow, and the system must be tested by a variety of end-users.
%, but the initial results are promising. 

%Note that the overall goal of the clinical trial was \textbf{not} a performance test. %We will have to perform better and more controlled experiments to properly gauge the performance of the model and the systems benefit in a clinical setting. 

\section{Discussion}\label{sec:discussion}
We have built a prototype for an end-to-end system for endoscopic video analysis for patients with UC. Our preliminary results indicate strong performance, although more clinical trials are needed to understand the true value of the proposed system. 
Our model achieves a Cohen's kappa of 0.79 on a hold-out test set of 254 images representing 84 patients not present in the training set, which suggests that it aligns well with the score of the committee of doctors that annotated the dataset. Notably, the model's alignment outperforms senior gastroenterologists~\cite{bad_MES_stats}. 
Furthermore, the model achieves an average AUROC of 0.904 when filtering out noisy data indicating it can handle noisy raw endoscopic video footage.   

Initially, our primary objective was to address the challenge of high inter-observer variability associated with the MES score. However, through close collaboration and ongoing discussions with clinicians, we have identified additional valuable use cases for the underlying machine learning model.
One important application we have discovered is the potential capability of the machine learning model to assist in frame selection during endoscopy procedures. By analyzing the real-time footage, the model can identify and highlight frames that are most informative or relevant to the clinician. This feature allows clinicians to focus their attention on key frames, potentially improving diagnostic accuracy and decision-making.
The decision about which frames should be selected and shown is complicated by the fact that the selected frames are also the frames that are candidates for relabeling, so there is a trade-off between increasing model performance and model explainability.  
Our current solution for selecting frames is to display the frames that affect the final video score the most. This choice was made to give the end-user as much information about the scoring as possible. The problem with this method is that important frames that are misclassified as being unimportant (false negatives) will never be shown. We are currently exploring options where one or more of the images displayed are not necessarily important for the overall score, but instead selected according to some active learning criterion. This should theoretically allow the model to be more robust towards some false negative images. %We have currently not found any active learning method, that outperforms random choice.

Furthermore, our machine learning model could also serve as a valuable tool in providing feedback to clinicians regarding the quality and suitability of the current footage. By indicating whether the current view is appropriate or if adjustments are required, the model can assist clinicians in positioning the endoscopic camera more effectively, ultimately leading to improved visualization and assessment during the procedure.

While not a requirement, it should be noted that our system allows for easy adaptation to different metrics. In case the MES score is replaced with another metric, such as the UECIS \cite{UECIS}, the system can be adjusted by relabeling the data accordingly. Our group is currently working on adapting the system to leverage the UECIS.
%This should allow our system to adapt to the changing recommendations of the medical field

Throughout the design process, we put emphasis on ensuring that the final system serves as a helpful assistant in a non-distracting manner. An interesting finding from our UX workshop is that clinicians were not particularly interested in real-time model performance metrics, as these metrics might unnecessarily distract the end-user. Future work should focus on exploring methods to provide model predictions in the least distracting manner, which in turn would enhance the usability and user experience of the system.

There are still design choices regarding model retraining which need to be explored. 
Our current idealized retraining protocol is that labelled data is anonymized and uploaded to a server, where a new model is retrained using all data. The new model is then redistributed from the server. This  allows us to extract data from many different clinics to generate one general model, which aligns the MES scoring across clinics. An alternative approach would be to refine a base model for each individual clinic based on local data. %is, that a model is trained for each clinic. Which method is best is currently an open question.  

The basic requirements outlined in \autoref{sec:req} are fulfilled by the current system design. The core model distinguished between all MES grades on both frame and video level operating on raw noisy video data. The system runs in real-time and can be connected to all endoscopic systems that have an SDI input and an output. The system can provide information about the quality of an endoscopic view. It will have a mechanism for improving over time by leveraging a combination of retraining and image relabeling. 
%Whether our current system can be improved to better solve some of these sub-goals require more experiments to fully grasp.  

Adopting medical decision support systems in clinical practice bears the risk of  end-user over-reliance.
Our system still requires the doctor to operate the camera with expertise and intent, it only provides feedback about whether the current view is good enough for scoring or not.
We envision our classification system to support clinicians by providing a second opinion on disease severeness and the preliminary results indicate that the model's classification accuracy is on par with real-world expert physicians \cite{Previous_work_on_dataset}, so over-reliance might not be a serious issue.
%However, 
%The potential future adoption of our system into the clinical environment might open the door to end-user overreliance. This issue is mitigated by two facts. Our preliminary results indicate that the model is on par with real-world expert physicians \cite{Previous_work_on_dataset}, so over-reliance might not be an issue. Furthermore, the system does not provide information about how to move the camera, only whether the current view is good enough for scoring. Thus the framework still requires the doctor to operate the camera with expertise and intent.

\section{Conclusion}\label{sec:conclusion}
We are developing the HECTOR system to provide automated assistance to clinicians during live endoscopy procedures. We have conducted our first clinical trial on a patient. Our preliminary results are promising, but much more clinical experimental data are needed for a  comprehensive assessment of the system's value and for refining the human-system interaction. % so it will integrate seamlessly into the daily clinical workflow.

\subsection*{Acknowledgements}
We acknowledge support from the Gastro Unit, medical section, Copenhagen University Hospital Hvidovre, which made this research possible. We thank the Novo Nordisk Foundation for aiding our research with financial support under grant NNF20OC0062056. Finally, we thank BETA.HEALTH for supporting our goal to bring our research into clinical practice under grant no 1260. 

\subsection*{Conflict Of Interest}
Bobby Lo reports an unrestricted grant from Beta.Health related to the submitted work and advisory/consultant fee from Tillotts Pharma, Bristol-Myers Squibb, Paratech A/S; Teaching fee from Tillotts Pharma, Janssen Cilag; Research grant from Tillotts Pharma, Janssen Cilag, Takeda, Abbvie.
Dr. Burisch reports grants and personal fees from AbbVie, grants and personal fees from Janssen-Cilag, personal fees from Celgene, grants and personal fees from MSD, personal fees from Pfizer, grants and personal fees from Takeda, grants and personal fees from Tillots Pharma, personal fees from Samsung Bioepis, grants and personal fees from Bristol Myers Squibb, grants from Novo Nordisk, personal fees from Pharmacosmos, personal fees from Ferring, personal fees from Galapagos,  outside the submitted work.
Doctor Bendtsen reports consulting fees, lecture fees or research funds from Takeda Pharma A/S, Norgine Danmark A/S, Ferring Pharmaceuticals A/S.
No other authors have anything to report.

{\footnotesize
\bibliography{ms}
}

\end{document}